\documentclass[iop,apjl]{emulateapj}
\newcommand\ksmpc{{~km~s$^{-1}$~Mpc$^{-1}$}\xspace}
\newcommand\lmass{{\log(M_\ast/M_\odot)}\xspace}

\usepackage{etoolbox}
\usepackage{xspace}
\usepackage{amsmath}
\usepackage{graphicx}
\usepackage[colorlinks,breaklinks, citecolor=blue,linkcolor=blue,urlcolor=blue]{hyperref}
\usepackage{apjfonts}
\makeatletter

\patchcmd{\NAT@citex}
    {\@citea\NAT@hyper@{\NAT@nmfmt{\NAT@nm}\NAT@date}}
    {\@citea\NAT@nmfmt{\NAT@nm}\NAT@hyper@{\NAT@date}}
    {}
    {}

\patchcmd{\NAT@citex}
    {\@citea\NAT@hyper@{%
         \NAT@nmfmt{\NAT@nm}%
         \hyper@natlinkbreak{\NAT@aysep\NAT@spacechar}{\@citeb\@extra@b@citeb}%
         \NAT@date}}
    {\@citea\NAT@nmfmt{\NAT@nm}%
     \NAT@aysep\NAT@spacechar%
     \NAT@hyper@{\NAT@date}}
    {}
    {}

\patchcmd{\NAT@citex}
    {\@citea\NAT@hyper@{%
         \NAT@nmfmt{\NAT@nm}%
         \hyper@natlinkbreak
         {\NAT@spacechar\NAT@@open\if*#1*\else#1\NAT@spacechar\fi}%
         {\@citeb\@extra@b@citeb}%
         \NAT@date}}
    {\@citea\NAT@nmfmt{\NAT@nm}%
        \NAT@spacechar\NAT@@open\if*#1*\else#1\NAT@spacechar\fi%
        \NAT@hyper@{\NAT@date}}
    {}
    {}
\makeatother

\slugcomment{Accepted to ApJ Letters}

\begin{document}

\shorttitle{Stellar populations of \emph{UVJ}-selected quiescent galaxies from KMOS}
\shortauthors{Mendel et al.}

\title{First results from the VIRIAL survey: the stellar content of \emph{UVJ}-selected quiescent galaxies at $1.5 < \MakeLowercase{z} < 2$ from KMOS\altaffilmark{$\dagger$}}

\author{J.\ Trevor Mendel\altaffilmark{1,2}, Roberto P. Saglia\altaffilmark{1,2}, Ralf Bender\altaffilmark{2,1}, Alessandra Beifiori\altaffilmark{2,1}, Jeffrey Chan\altaffilmark{1,2}, Matteo Fossati\altaffilmark{2,1}, David J. Wilman\altaffilmark{2,1}, Kaushala Bandara\altaffilmark{1}, Gabriel B. Brammer\altaffilmark{3}, Natascha M. F\"orster Schreiber\altaffilmark{1}, Audrey Galametz\altaffilmark{1,2}, Sandesh Kulkarni\altaffilmark{1,2}, Ivelina G. Momcheva\altaffilmark{4}, Erica J. Nelson\altaffilmark{4}, Pieter G. van Dokkum\altaffilmark{4},\\ Katherine E. Whitaker\altaffilmark{5}, and Stijn Wuyts\altaffilmark{1}}

\affil{\altaffilmark{1}{Max-Planck-Institut f\"{u}r Extraterrestrische Physik, Giessenbachstr. 1, 85748 Garching, Germany; \href{mailto:jtmendel@mpe.mpg.de}{jtmendel@mpe.mpg.de}}\\
\altaffilmark{2}{Universit\"{a}ts-Sternwarte M\"{u}nchen, Scheinerstr. 1, 81679 M\"{u}nchen, Germany}\\
\altaffilmark{3}{Space Telescope Science Institute, Baltimore, MD 21218, USA}\\
\altaffilmark{4}{Department of Astronomy, Yale University, New Haven. CT 06511, USA}\\
\altaffilmark{5}{Astrophysics Science Division, Goddard Space Flight Center, Code 665, Greenbelt, MD 20771, USA}
}

\altaffiltext{$\dagger$}
{Based on observations obtained at the Very Large Telescope (VLT) of the European Southern Observatory (ESO), Paranal, Chile (ESO program IDs 092.A-0091, 093.A-0079, 093.A-0187, and 094.A-0287).  This work is further based on observations taken by the 3D-HST Treasury Program (GO 12177 and 12328) with the NASA/ESA \emph{Hubble Space Telescope}, which is operated by the Association of Universities for Research in Astronomy, Inc., under NASA contract NAS5-26555.}

\begin{abstract}

We investigate the stellar populations of 25 massive, galaxies ($\log[M_\ast/M_\odot] \geq 10.9$) at $1.5 < z < 2$ using data obtained with the \emph{K}-band Multi-Object Spectrograph (KMOS) on the ESO VLT.  Targets were selected to be quiescent based on their broadband colors and redshifts using data from the 3D-HST grism survey.  The mean redshift of our sample is $\bar{z} = 1.75$, where KMOS \emph{YJ}-band data probe age- and metallicity-sensitive absorption features in the rest-frame optical, including the $G$ band, Fe I, and high-order Balmer lines.  Fitting simple stellar population models to a stack of our KMOS spectra, we derive a mean age of $1.03^{+0.13}_{-0.08}$~Gyr.  We confirm previous results suggesting a correlation between color and age for quiescent galaxies, finding mean ages of $1.22^{+0.56}_{-0.19}$ Gyr and $0.85^{+0.08}_{-0.05}$ Gyr for the reddest and bluest galaxies in our sample.  Combining our KMOS measurements with those obtained from previous studies at $0.2 < z < 2$ we find evidence for a $2-3$ Gyr spread in the formation epoch of massive galaxies.  At $z < 1$ the measured stellar ages are consistent with passive evolution, while at $1 < z \lesssim2$ they appear to saturate at $\sim$1 Gyr, which likely reflects changing demographics of the (mean) progenitor population.  By comparing to star-formation histories inferred for ``normal'' star-forming galaxies, we show that the timescales required to form massive galaxies at $z \gtrsim 1.5$ are consistent with the enhanced $\alpha$-element abundances found in massive local early-type galaxies.

\end{abstract}

\keywords{galaxies: evolution --- galaxies: formation --- galaxies: high-redshift }

\section{Introduction}
\label{intro}

At early times, galaxies' growth is dominated by in-situ star-formation fed primarily by direct accretion of gas from the cosmic web \citep[e.g.,][]{dekel2009,tacconi2010}.  The relative balance of accretion, star formation, and feedback in this ``equilibrium growth'' phase leads to a tight relationship between star-formation rate (SFR) and stellar mass for normal star-forming galaxies \citep[e.g.,][]{bower2006,schaye2015}.  The overall normalization of this star-forming ``main sequence'' evolves downward from the epoch of peak activity at $z \sim 2$ to the present \citep[e.g.,][]{daddi2007,rodighiero2011,whitaker2014,schreiber2015}, reflecting both a declining accretion rate of gas onto halos and the steady depletion of gas via star formation \citep[e.g.,][]{lilly2013}.  Galaxies appear to spend the majority of their lifetimes on the main sequence \citep[e.g.,][]{noeske2007} until their star formation is ``quenched'' near the Schechter mass \citep[e.g.,][]{peng2010}.  This quenching process leads to formation of a quiescent galaxy population whose subsequent evolution is dominated by the assembly of already-formed stellar mass via mergers \citep[e.g.,][]{oser2010,moster2013}.

In the local Universe, quiescent galaxies occupy a well-defined ``red sequence'' in terms of their color and luminosity (or stellar mass) with a very small intrinsic scatter \citep[$\sigma<0.04$~mag; e.g.,][]{bower1992a} suggesting an early, rapid formation process which is supported by their old ages and high $\alpha$-element abundance ratios \citep[e.g.,][]{thomas2005}.  Over the last decade, near-infrared photometric surveys have shown that the massive end of the red sequence is in place already by at least $z \sim 2$ \citep[e.g.,][]{williams2009,brammer2009}, consistent with the early formation epochs inferred from local elliptical galaxies; however, the details of their formation remain poorly understood, motivating a more direct study of their stellar populations, and in particular ages and $\alpha$-element abundances, which relate directly to their star-formation histories.  Only recently--with the development of efficient red-sensitive optical and near-infrared detectors--has it been possible to measure galaxy stellar populations routinely up to $z \sim 1$ \citep[e.g.][]{sanchez-blazquez2009,belli2015} and as high as $z \sim 2$ for select sub-samples \citep{belli2014,onodera2014}.

In this \emph{Letter} we investigate the stellar populations of massive quiescent galaxies at $1.5 < z < 2$ using new data obtained with KMOS \citep{sharples2012,sharples2013} as part of the VLT IR IFU Absorption Line GTO survey (VIRIAL; Mendel et al. in prep.).  These data probe rest-frame \emph{V}-band absorption features for galaxies at $z > 1.5$ with a resolution of $R\sim 3500$, allowing us to study the stellar content of massive galaxies near the epoch when they are being formed.  

Where quoted, magnitudes are on the AB system, and stellar masses have been computed using a \citet{chabrier2003} stellar initial mass function (IMF).  We adopt a flat cosmology with $\Omega_{\Lambda} = 0.7$, $\Omega_{\mathrm{M}} = 0.3$ and $H_0 = 70$\ksmpc.

\section{Data and analysis}
\label{data}

\begin{figure}
\centering
\includegraphics[scale=0.75]{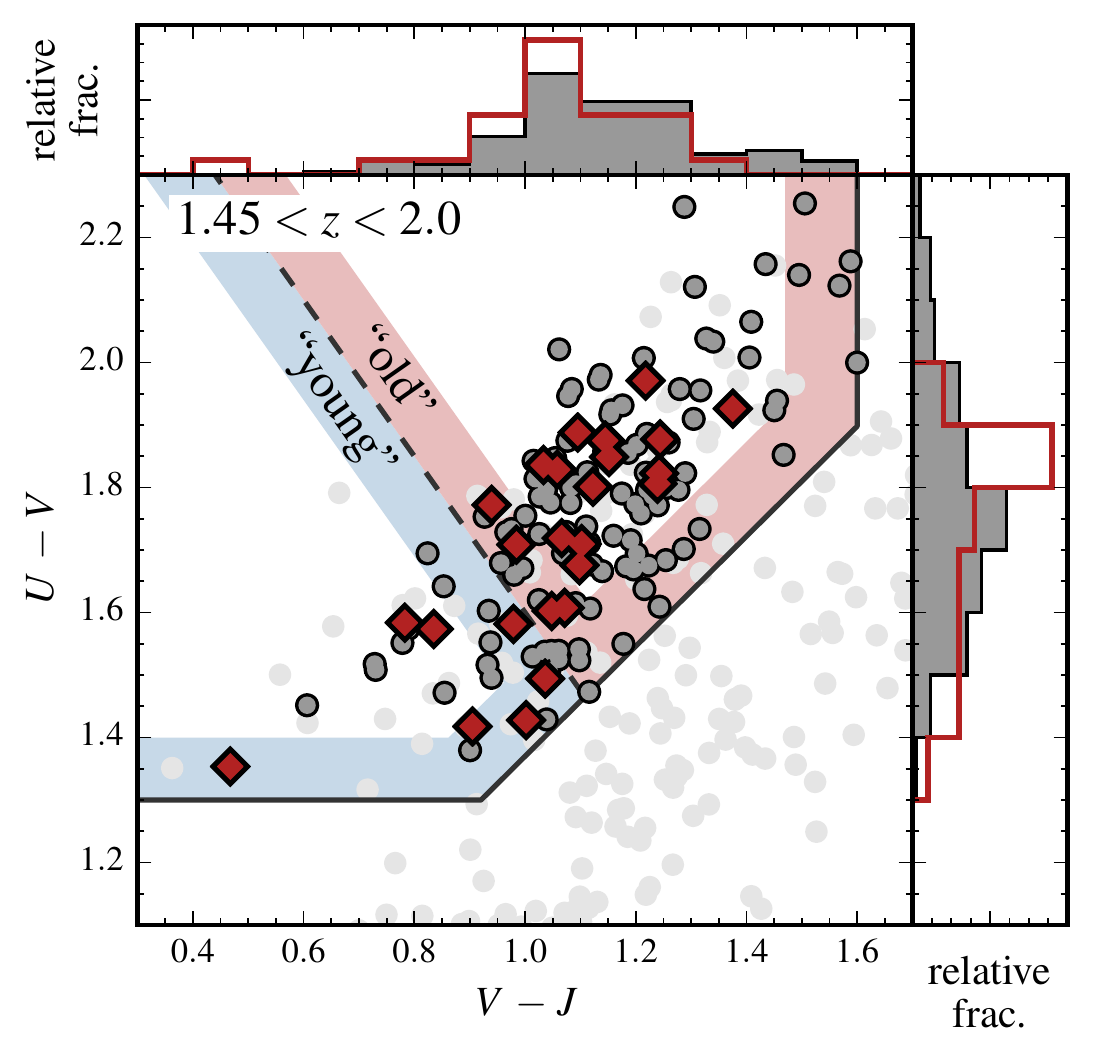}
\caption{VIRIAL survey sample selection.  Light grey points show the distribution of 3D-HST targets at $1.45 < z < 2$ while darker circles highlight those galaxies satisfying the selection criteria described in Section \ref{data}.  Diamonds indicate galaxies considered in the present work.  The division between nominally ``old'' and ``young'' galaxies based on their colors is adopted from \citet{whitaker2013}.  The inset histograms show the normalized color distribution for the full sample (filled) and the sub-sample used here (open).  The current data are roughly representative of the underlying $UVJ$-quiescent population.}
\label{fig:uvj}
\end{figure}

VIRIAL survey targets are selected based on combined redshift, magnitude, and color criteria using data from the 3D-HST Treasury Program \citep{brammer2012,skelton2014} and CANDELS \citep{grogin2011,koekemoer2011} in the COSMOS, GOODS-S, and UDS fields.  We first identify quiescent galaxies in the redshift range $1.45 < z < 2$ by their rest-frame $U\!-\!V$ and $V\!-\!J$ colors (hereafter $UVJ$; Figure \ref{fig:uvj}) using the division described by \citet[see also \citealp{whitaker2011}]{williams2009}, supplementing publicly-available spectroscopic redshifts with those derived using 3D-HST grism and photometric data.  We then consider only those galaxies with $m_\mathrm{F140W} \leq 22.5$, corresponding to a sample which is $>$90\% complete for $\lmass \geq10.9$.  The adopted selection includes $>$95\% of the quiescent galaxies\footnote{Where quiescent galaxies are defined as those with SFRs a factor of 10 or more below the mean SFR for their stellar mass and redshift given by \citet{whitaker2014}.} brighter than our adopted magnitude limit.  The full VIRIAL survey consists of 132 galaxies which satisfy the above cuts. In this letter we focus on a sub-sample of the spectroscopic data with reliable redshifts obtained during the first 2 semesters of observations.  These galaxies have a color distribution which is roughly representative of the underlying target population, as shown by the histograms in Figure \ref{fig:uvj}.

\subsection{Observations and data reduction}
\label{obs}

\begin{figure*}
\centering
\includegraphics[scale=0.85]{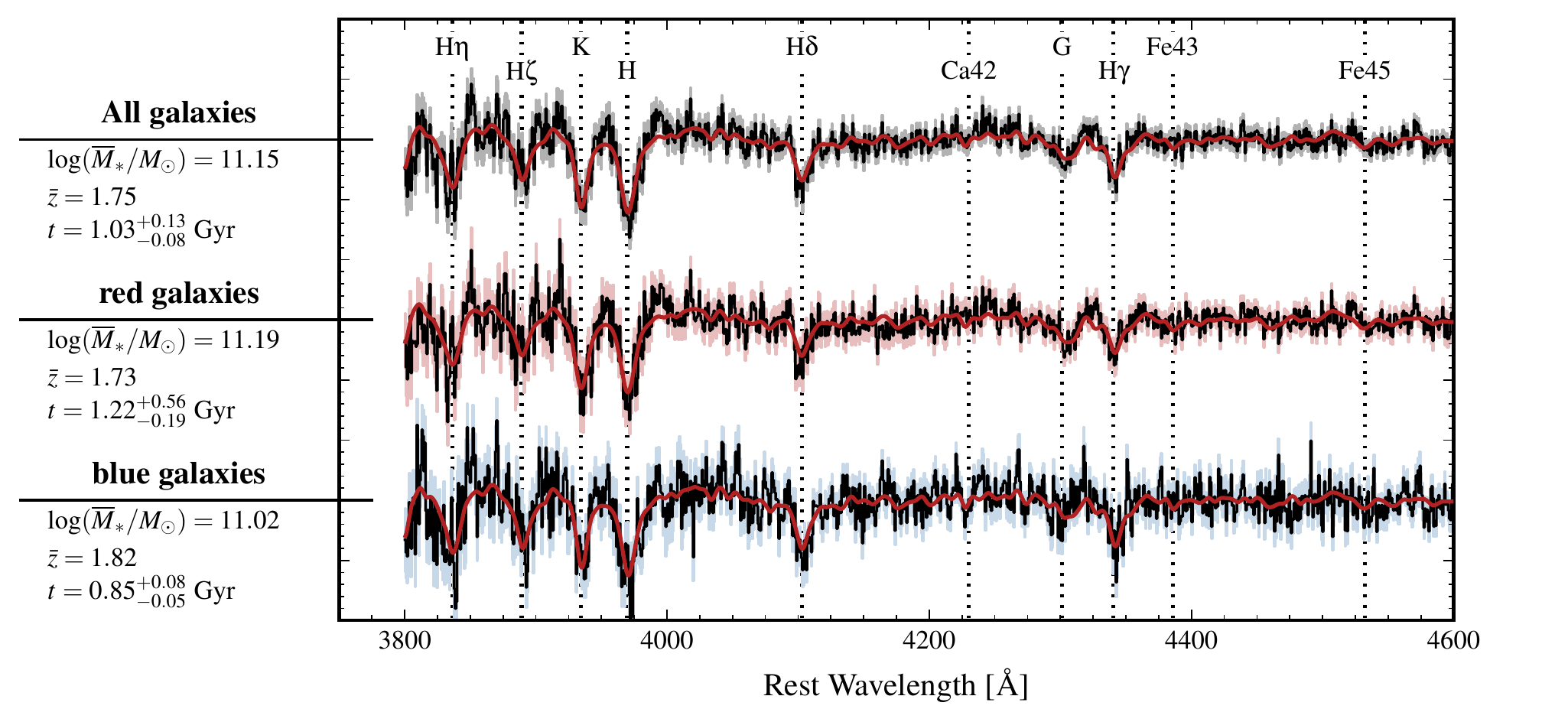}
\caption{Stacked galaxy spectra and their best-fit SSP models for the total (top), red (middle), and blue (bottom) subsamples. Spectra have been normalized by a high-order polynomial, and are offset for clarity.  Key absorption features are indicated by vertical lines.  Shading shows the $\pm1\sigma$ uncertainties on the stacked spectra, derived using bootstrap samples.}
\label{fig:spec_fig}
\end{figure*}

Observations of the selected VIRIAL targets were obtained on 2014 April 6-7, August 18-20, 28-30, September 1, and November 27-30.  Data were taken with a standard object-sky-object nodding pattern so that each on-source frame has an adjacent sky exposure.  We included small pointing offsets of $0.1-0.6^{\prime\prime}$ in each exposure to avoid an undue contribution from bad pixels in the final extracted spectra.  Typical exposure times were 300s in the \emph{YJ} band (1-1.36$\mu m$), and the total integration time per target ranges from 6-11 hours on source.  

Data were reduced using a combination of the Software Package for Astronomical Reductions with KMOS pipeline tools \citep[\texttt{SPARK};][]{davies2013} and custom {\tt Python} scripts.  Dark, arc, and flat frames were processed using standard {\tt SPARK} routines.  For both object and sky frames, we applied a correction for the channel-dependent bias level drift determined using reference pixels on the perimeter of each detector prior to reconstructing the data cubes.  Cubes were then corrected for spatial illumination effects using the observed sky-line fluxes on a frame-by-frame basis.  Sky subtraction was carried out using the methods described by \citet{davies2007} to correct for short-term variability in the OH airglow lines.  Finally, we corrected for telluric absorption using synthetic atmospheric models computed using {\tt MOLECFIT} \citep{kausch2014}.  Observations of spectrophotometric standards taken on each night were used to rescale the precipitable water vapor predicted by {\tt MOLECFIT} in order to better match observed conditions.  We then computed telluric corrections for each exposure using the predicted time-dependent model atmosphere and this derived scaling. 

In order to minimize systematic uncertainties in the inter-line sky background, one-dimensional spectra for each object were extracted using a routine which iteratively fits for an exposure- and wavelength-dependent background term contemporaneously with extraction of the source spectrum.  Since our targets are generally undetected in individual KMOS exposures we use the CANDELS/3D-HST F125W image mosaics\footnote{\url{http://http://3dhst.research.yale.edu/Data.php}} \citep{grogin2011,koekemoer2011,skelton2014} as a model for source flux distribution, which were convolved to match the KMOS PSF determined from reference stars assigned to IFUs in each exposure.  Uncertainties on the extracted spectra were estimated using 50 bootstrap realizations of the input frames.

\subsection{Stellar population measurements}
\label{sp_meas}

The signal-to-noise (S/N) of our extracted KMOS spectra is generally too low to yield robust stellar population measurements for individual galaxies; however, we can estimate the mean properties for galaxies in our sample by stacking their spectra given a sufficiently accurate redshift.

We estimated redshifts by first cross-correlating the extracted spectra with simple stellar population (SSP) templates at a variety of ages.  In the current work we limit our sample to the 25 galaxies with a clear peak in the cross-correlation signal.  By requiring well-measured redshifts we impose a bias in our sample towards low mass-to-light ratios;  however, the selected galaxies are nevertheless representative of the underlying $UVJ$-selected population (see Figure \ref{fig:uvj}).  There is generally excellent agreement between our KMOS-derived redshifts and those in the 3D-HST target catalog, with a median absolute deviation of $\sim$1300 km~s$^{-1}$, consistent with the uncertainties estimated by \citet{whitaker2013}.  Individual spectra were then normalized using a high-order polynomial before being co-added using inverse-variance weights.  In total we generated 3 stacks: one for the full sample, and two more where we divide the sample into red and blue sub-populations based on their $UVJ$ colors as in \citet{whitaker2013}.  The final stacks have S/Ns of 16~\AA$^{-1}$, 13~\AA$^{-1}$, and 9~\AA$^{-1}$ and are shown in Figure \ref{fig:spec_fig}. 

We estimated the properties of our stacked spectra using SSP models constructed with v2.4 of the Flexible Stellar Population Synthesis ({\tt FSPS}) code described by \citet{conroy2009}.  We fit simultaneously for the SSP-equivalent age $t$, total metallicity, and parameters of line-of-sight velocity distribution, using the ensemble sampling code {\tt emcee} \citep{foreman-mackey2013}. SSP models were linearly interpolated in $\log\,t$ and metallicity on the fly as part of the fitting procedure.  Fits were limited to $\lambda$3800--4600\AA~in order to exclude wavelength regions where only a few galaxies contribute.

Final measurements of the stellar age were taken as the median of the marginal posterior distribution; uncertainties were estimated using 100 bootstrap realizations of the stacked spectra, recomputing the best-fit age for each using the procedure outlined above.  We derive a mean age for the full stack of $1.03^{+0.13}_{-0.08}$~Gyr, while for red and blue sub-samples we derive ages of $1.22^{+0.56}_{-0.19}$~Gyr and $0.85^{+0.08}_{-0.05}$~Gyr.  In all cases the data are consistent with solar metallicity.  We stress that these uncertainties do not include systematic effects due to our choice of SSP model.

\section{Results}

\subsection{Quiescent galaxy stellar populations at $\lowercase{z} \approx 1.75$}

The relatively young mean age of massive galaxies in our sample suggests that most are recent additions to the quiescent population, consistent with other studies that have reported a large fraction of post-starburst-like galaxies at high redshift \citep[e.g.,][]{whitaker2012,bezanson2013}.  However, the spread in rest-frame colors among quiescent galaxies is likely driven at least in part by a variation in their formation times or star-formation histories, such that the reddest galaxies are also the oldest \citep{whitaker2010,whitaker2013}.  

\begin{figure}
\centering
\includegraphics[scale=0.82]{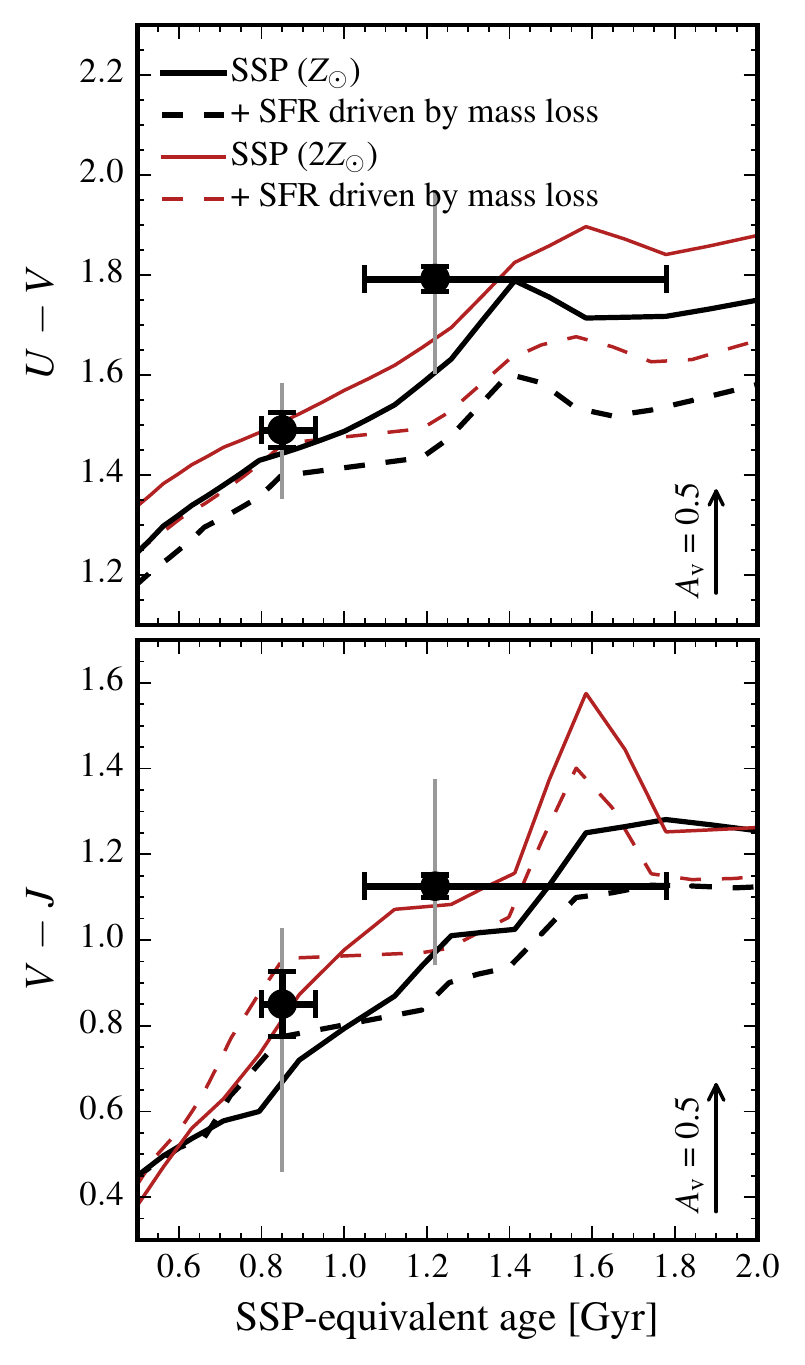}
\caption{$U\!-\!V$ and $V\!-\!J$ color as a function of stellar age for the red and blue galaxy populations.  Lines show predicted color vs.~age for SSP models at two different metallicities (solid lines), as well for models which include star-formation driven by stellar mass loss (dashed lines).  Uncertainties show the error on the mean derived from bootstrap samples, while thin lines indicate the range of colors present in each sample.  Vertical arrows indicate the expected change in color for $A_\mathrm{v} = 0.5$.  The measured ages are consistent with expectations from a passively-evolving SED model.}
\label{fig:color}
\end{figure}

In Figure \ref{fig:color} we show the color vs.~age derived from our KMOS data when they are split into two different bins of rest-frame color; the derived ages--as well as the $\sim$0.4 Gyr age difference between the two samples--are consistent with the evolution of a passively-evolving stellar population (solid lines in Figure \ref{fig:color}).  For comparison, we also consider SSP models which host residual star-formation fed by stellar mass loss, assuming that 100\% of the material lost is recycled into stars (dashed lines in Figure \ref{fig:color}).  While the bluest galaxies are compatible with either model, the reddest are unlikely to host significant ongoing star-formation, more than a few $M_\odot~\mathrm{yr^{-1}}$, unless they are also preferentially dusty.  Of the 25 galaxies in our sample, 4 (16\%) are detected at $24~\mu m$ in the available \emph{Spitzer}-MIPS data, suggesting that they may host obscured star formation; however, there is no clear preference for $24~\mu m$-detected sources to fall in a particular region of the $UVJ$-quiescent selection window \citep[see also][]{fumagalli2014}.  

\subsection{Mean evolution of red galaxies from $\lowercase{z} \approx 2$}
\label{age_evo}

\begin{figure}
\centering
\includegraphics[scale=0.62]{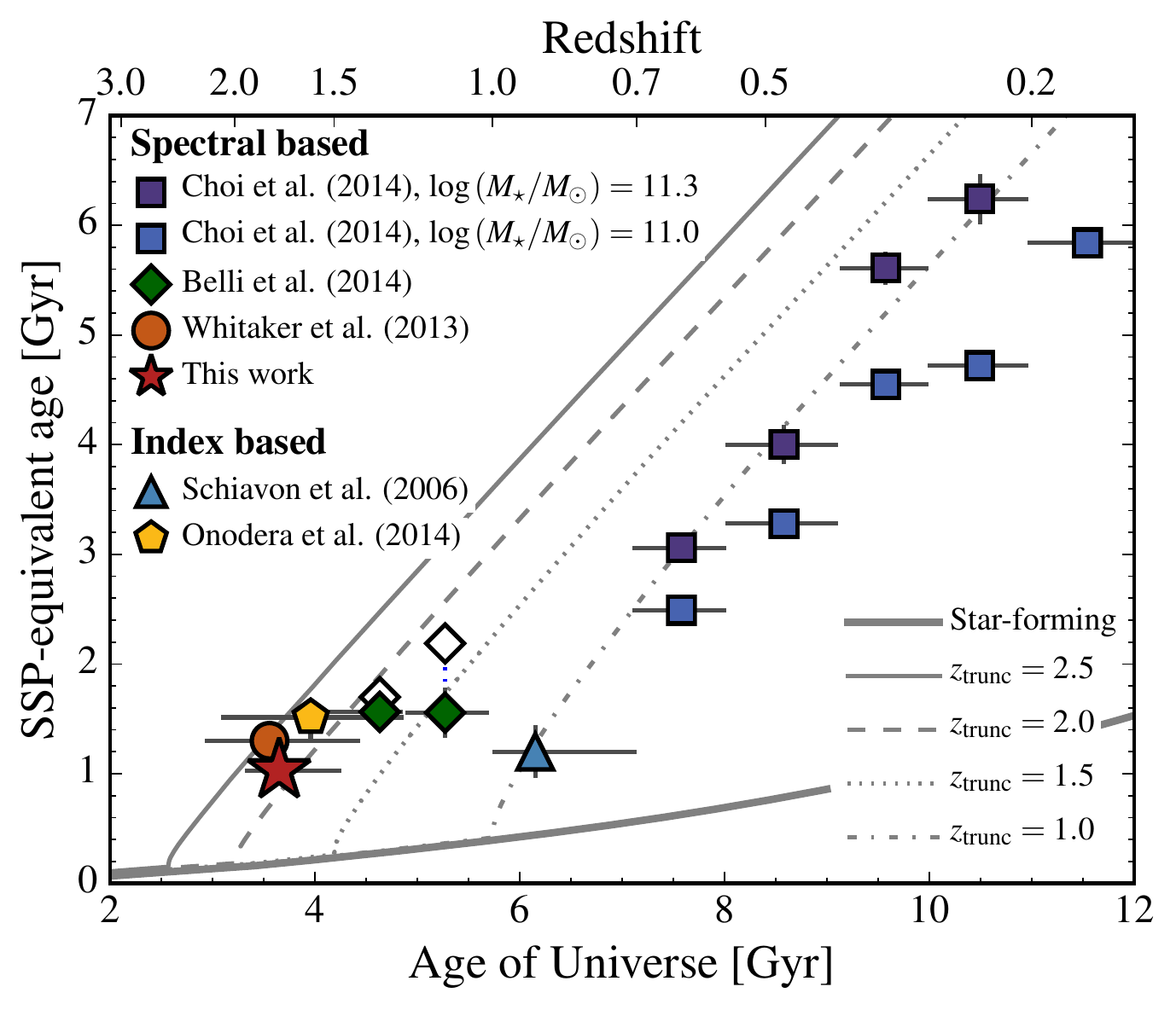}
\caption{Evolution of mean stellar age for massive quiescent galaxies over the past $\sim$9 Gyr.  Our age estimates, as well as those from \citet{schiavon2006}, \citet{onodera2014}, \citet{whitaker2013}, and \citet{choi2014}, are measured from stacked data, while the points from \citet{belli2015} are taken as the median of their individually-analyzed galaxy spectra split into two redshift bins.  We have corrected the \citet{belli2015} values from their $\tau$-model parametrization to light-weighted values for comparison with the other data (original values are shown as open diamonds).  Error bars indicate either the uncertainty in a given SSP-equivalent age measurement, or range in redshift spanned by a given sample.  Lines show evolution of the light-weighted age for the truncated MS star-formation histories described in Section \ref{age_evo}.  The flattening of mean stellar ages at $z > 1$ likely reflects the increased importance of quenching to the formation of massive quiescent galaxies at high redshift.}
\label{fig:age_evo}
\end{figure}

In Figure \ref{fig:age_evo} we investigate evolution of galaxy ages from $z \approx 2$ to the present, incorporating recent results from the literature.  We include age estimates based on spectral fitting from \citet{choi2014} at $0.1 < z < 0.7$, \citet{belli2015} at $1 < z < 1.5$, and \citet{whitaker2013} at $1.4 < z < 2.2$.  The \citet{belli2015} data have been converted from their $\tau$-model parameterization to light-weighted values using the age, $\tau$, and metallicity as given in their Table 1; we plot here the median age and redshift of their data split into two redshift bins.  Two additional age estimates based on fitting to standard Lick absorption line indices \citep[e.g.][]{worthey1994} from \citet{schiavon2006} at $\bar{z} = 0.9$ and \citet{onodera2014} at $\bar{z} = 1.6$ are also shown.  

As a comparison to the observed ages we consider the evolution of galaxies which form as a part of the normal star-forming population, but whose star formation is truncated after some redshift $z_\mathrm{trunc}$ so that they evolve passively to the present day.  We reconstruct mean star-formation histories (SFHs) by considering that the evolution of stellar mass is driven by a combination of star formation and stellar mass loss \citep[e.g., ][]{leitner2011,leitner2012}, i.e.

\begin{equation}
\dot{M}_\ast(t) = \mathrm{SFR}(M_\ast,z) - \Re (t),
\label{growth}
\end{equation}

\noindent where $\mathrm{SFR}(M_\ast,z)$ is the redshift- and mass-dependent SFR, and $\Re(t)$ is a convolution of the SFH and stellar mass-loss rate, which we estimate using {\tt FSPS}.  Following \citet{lilly2013} we assume that the typical specific SFR (sSFR $\equiv$ SFR / $M_\ast$) of a star-forming galaxy with mass $M_\ast$ at redshift $z$ is

\begin{equation}
\mathrm{sSFR}(M_{\ast},z) = \left\{\!\begin{array}{ll}
0.07\left( \frac{M_{\ast}}{10^{10.5}M_{\odot}} \right)^{-0.1} (1+z)^{3} & \mbox{if $z\leq2$}\\
0.30\left( \frac{M_{\star}}{10^{10.5}M_{\odot}} \right)^{-0.1} (1+z)^{5/3}  & \mbox{if $z>2$}.
\end{array}\right.
\label{ssfr}
\end{equation}

\noindent For a given final stellar mass, $M_\ast(z_\mathrm{trunc})$, we evolve galaxies back in time according to Equation \ref{growth}, using the iterative procedure described by \citet{leitner2011} to solve for $\dot{M}_\ast(t)$, until they reach an initial mass of $10^8~M_\sun$; the remaining mass is assumed to form in a burst at $z = 8$.  The resulting main-sequence lifetime--from an initial mass of $10^8~M_\sun$ to $M_\ast(z_\mathrm{trunc})$--is only weakly dependent on the particular value of the initial mass, for example adopting $10^{7}~M_\sun$ results in lifetimes which are longer by only $\sim150$ Myr.  In Figure \ref{fig:age_evo} we show the evolution of luminosity-weighted age for SFHs with $1 \leq z_\mathrm{trunc} \leq 2.5$ and $M_\ast(z_\mathrm{trunc}) = 10^{11} M_\odot$, as well as for a galaxy which remains star forming until the present day (i.e. $z_\mathrm{trunc} = 0$).  

There are two competing effects which drive the age evolution shown in Figure \ref{fig:age_evo}: the quenching of star formation in star-forming galaxies, and the passive evolution of already-formed stellar mass.  At $z < 1$ the stellar populations of massive galaxies are consistent with simple passive evolution, in agreement with previous spectroscopic studies \citep[e.g.,][]{choi2014}, as well as with evolution of the galaxy luminosity function over a similar epoch \citep[e.g.,][]{brown2007}.  On the other hand, the relatively uniform ages measured at $z > 1$ suggest that the quiescent galaxy population is being kept young (in the mean) by the constant addition of recently-quenched galaxies.  The transition at $z\approx1$ therefore reflects changing demographics of the progenitor population \citep[i.e., progenitor bias;][]{van-dokkum2001}, where the factor of $\sim$3 increase in quiescent galaxy number density from $z = 2$ to 1 \citep[e.g.,][]{tomczak2014} is dominated by the addition of recently-quenched star-forming galaxies, and at $z < 1$ the massive galaxy population grows primarily by the assembly of existing quiescent galaxies.  \citet{marchesini2014} obtain qualitatively similar results for the most massive galaxies by tracing progenitors at fixed number density \citep[see also][]{papovich2014}.  Based on comparison with the mock main-sequence SFHs, a $\sim2-3$ Gyr spread in formation times is required to explain both the relatively old ages found for quiescent galaxies already at $z = 1.5-2$ as well as the ages measured at $z=0$, but is nonetheless consistent with the small scatter observed locally among red-sequence galaxies \citep[e.g.,][]{bower1992a}.

\subsection{The formation timescale of quiescent galaxies}

\begin{figure}
\centering
\includegraphics[scale=0.61]{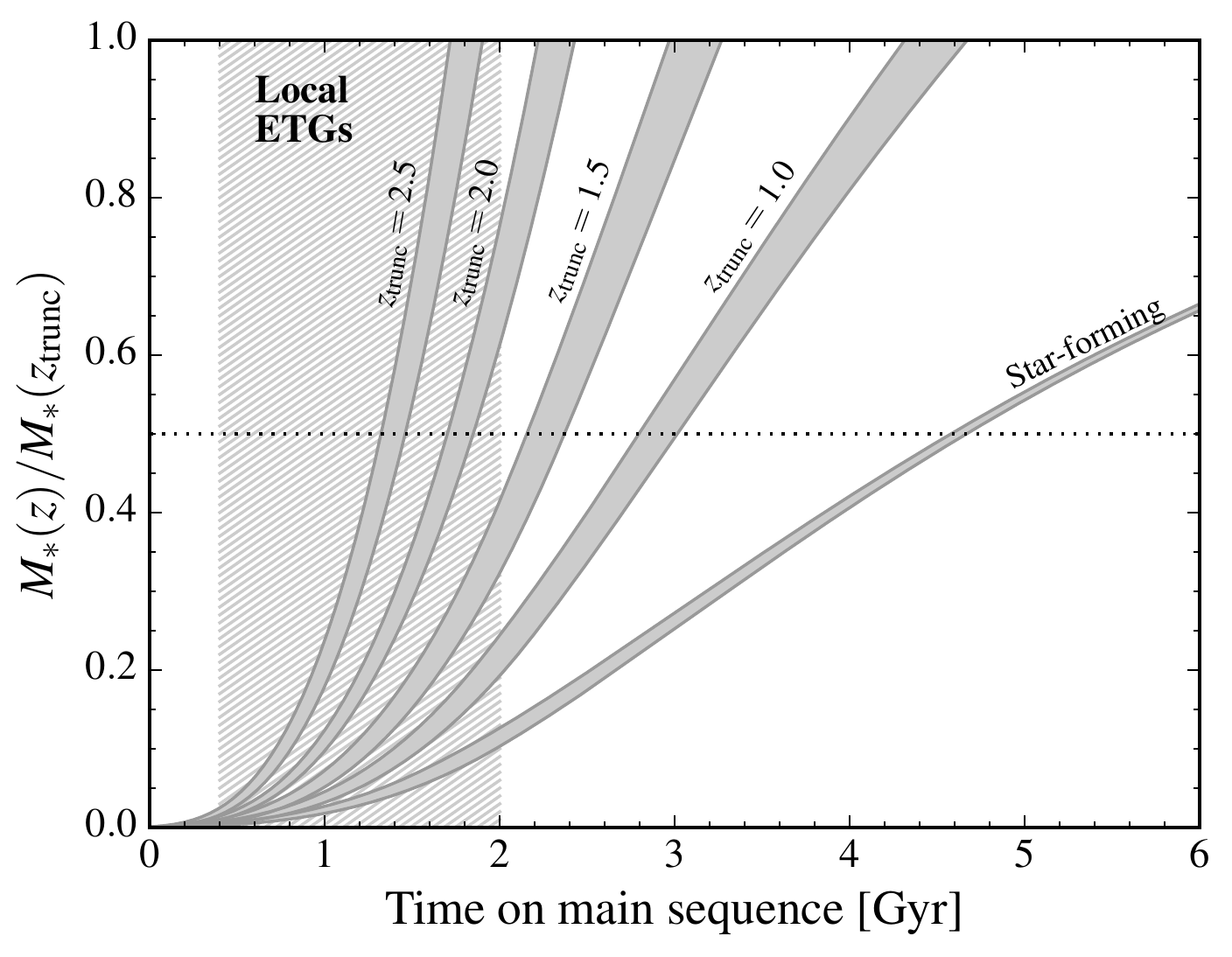}
\caption{Stellar mass growth as a function star-formation timescale for the truncated star-formation histories described in Section \ref{age_evo}.  Shaded regions trace the mass buildup for galaxies with $10^{11} M_\odot \leq M_\ast(z_\mathrm{trunc}) \leq 10^{11.5} M_\odot$ after they join the main sequence population.  The point where galaxies have formed 50\% of their final mass is marked by the dotted line.  The vertical hatched region shows the range of formation timescales for local early-type galaxies implied by their (enhanced) $\alpha$-element abundances.  The high star-formation rates at $z \gtrsim 2$ imply that massive quiescent galaxies can form as part of the normal star-forming population on very short timescales.}
\label{fig:ms_evo}
\end{figure}

In the previous Section we considered possible evolutionary tracks for quiescent galaxies assuming that their progenitors evolve as normal star-forming galaxies.  In Figure \ref{fig:ms_evo} we show the implied mass growth for these models as a function of star-formation timescale, in this case defined as the time galaxies spend on the main sequence starting from an initial mass of $10^8 M_\odot$.

The chemical abundance patterns of local ellipticals suggest that they form on relatively short timescales, $0.4-2$~Gyr \citep[e.g.,][shown as the hatched region in Figure \ref{fig:ms_evo}]{thomas2005,johansson2012}, which are consistent with the main-sequence lifetimes inferred for massive quiescent galaxies at $z \gtrsim 1.5$.  On the other hand, the rapid evolution of sSFR $\propto (1+z)^3$ at $z < 2$ (Equation \ref{ssfr}) leads to a dramatic increase in the formation time of massive galaxies on the main sequence, so that by $z \approx 1$ the typical timescale is a factor of $\sim$2 longer than that inferred from the $\alpha$-element abundances of local early-types.  It is important to note that this comparison neglects the role of mergers: \citet{oser2010} show that massive galaxies can accrete $\sim$50\% of their final stellar mass from $z = 1$ to 0.  Therefore, at least some of the tension between main-sequence lifetime and formation timescale implied by Figure \ref{fig:ms_evo} for galaxies at $z < 1$ can be alleviated by assuming that the majority of their stars formed in lower-mass galaxies, which then assemble at late times.

\section{Summary}

We present new estimates for the ages of massive quiescent galaxies derived from deep KMOS observations.  These data were obtained as part of the on-going VIRIAL GTO program, which targets quiescent galaxies selected from the 3D-HST survey.  By stacking the current data we estimate a mean luminosity-weighted age of $1.03^{+0.13}_{-0.08}$ Gyr for quiescent galaxies at $z \approx 1.75$.  Separating galaxies based on their $UVJ$ colors, we confirm the correlation between color and luminosity-weighted age among quiescent galaxies, deriving mean ages of $1.22^{+0.56}_{-0.19}$ Gyr and $0.85^{+0.08}_{-0.05}$ Gyr for the red and blue subpopulations.  The measured age and color differences are consistent with a passively-evolving SSP, suggesting only minor contribution from ongoing star-formation.  Investigating the evolution of galaxy ages from $z \approx 2$ to 0, we show that the current data suggest a spread in formation times of at least $2-3$ Gyr in order to explain both the mean ages at $z > 1$ and passive evolution at $z < 1$.  There is a clear transition in the behavior of galaxy ages at $z \approx 1$ which likely reflects the changing properties of the progenitor population.  Based on a comparison to the star-formation histories of main sequence galaxies at different redshifts, we show that the formation timescale inferred for massive galaxies at $z \gtrsim 1.5$ is consistent with the enhanced $\alpha$-element abundances observed in local ETGs, suggesting that such systems may plausibly form as part of the ``normal'' star-forming population.

\acknowledgements

We are indebted to the entire KMOS instrument and commissioning teams for their hard work, which has allowed our observing program to be carried out successfully.  We wish to thank the ESO staff, and in particular the staff at Paranal Observatory, for their support during observing runs over which the KMOS GTO observations were carried out.  We also thank the anonymous referee for their constructive comments.  DJW and MF acknowledge the support of the Deutsche Forschungsgemeinschaft via Project ID 387/1-1.

\bibliographystyle{apj}

\end{document}